\documentclass[twocolumn,showpacs,preprintnumbers,amsmath,amssymb,superscriptaddress,prl]{revtex4}

\usepackage{graphicx}
\usepackage{dcolumn}
\usepackage{bm}

\begin{document}

\hyphenation{te-tra-go-nal}

\bibliographystyle{apsrev}

\title{Experimental proof of a structural origin for the shadow Fermi surface in Bi$_2$Sr$_2$CaCu$_2$O$_{8+\delta}$}

\author{A. Mans}
\affiliation{Van der Waals-Zeeman Institute, University of
Amsterdam, NL-1018XE Amsterdam, The Netherlands}
\author{I. Santoso}
\affiliation{Van der Waals-Zeeman Institute, University of
Amsterdam, NL-1018XE Amsterdam, The Netherlands}
\author{Y. Huang}
\affiliation{Van der Waals-Zeeman Institute, University of
Amsterdam, NL-1018XE Amsterdam, The Netherlands}
\author{W. K. Siu}
\affiliation{Van der Waals-Zeeman Institute, University of
Amsterdam, NL-1018XE Amsterdam, The Netherlands}
\author{S. Tavaddod}
\affiliation{Van der Waals-Zeeman Institute, University of
Amsterdam, NL-1018XE Amsterdam, The Netherlands}
\author{V. Arpiainen}
\affiliation{Tampere University of Technology, Department of
Physics, PO Box 692, FIN-33101 Tampere, Finland}
\author{M. Lindroos}
\affiliation{Tampere University of Technology, Department of
Physics, PO Box 692, FIN-33101 Tampere, Finland}
\author{H. Berger}
\affiliation{Ecole Polytechnique F\'{e}d\'{e}rale de Lausanne, Institut de Physique de la Mati\`{e}re Complexe
EPFL Bât. PH CH-1015}
\author{V. N. Strocov}
\affiliation{Swiss Light Source, Paul Scherrer Institute, CH-5232 Villigen, Switzerland}
\author{M. Shi}
\affiliation{Swiss Light Source, Paul Scherrer Institute, CH-5232 Villigen, Switzerland}
\author{L. Patthey}
\affiliation{Swiss Light Source, Paul Scherrer Institute, CH-5232 Villigen, Switzerland}
\author{M. S. Golden}
\affiliation{Van der Waals-Zeeman Institute, University of Amsterdam, NL-1018XE Amsterdam, The Netherlands}

\begin{abstract}

By combining surprising new results from a full polarization analysis of nodal angle-resolved photoemission data
from pristine and modulation-free Bi$_2$Sr$_2$CaCu$_2$O$_{8+\delta}$ with structural information from LEED and
{\it ab initio} one-step photoemission simulations, we prove that the shadow Fermi surface in these systems has
structural origin, being due to orthorhombic distortions from tetragonal symmetry present in both surface and
bulk. Consequently, one of the longest standing open issues in the fermiology of these widely studied systems
finally meets its resolution.
\\

\end{abstract}

\maketitle

The Fermi surface and associated low-lying electronic excitations are central to the physical properties of metallic
solids. Thus in systems displaying highly anomalous and complex electronic behavior, such as the high-$T_c$ cuprate
superconductors, much effort is expended in the study of their 'fermiology' using photoemission (ARPES) \cite{reviews}.
In this context, although Bi$_2$Sr$_2$CaCu$_2$O$_{8+\delta}$ (Bi2212) is one of the most intensively studied solids
known, significant pieces of the Fermi surface puzzle for these compounds have still managed to remain in the shadows.
Fittingly, it is the so-called shadow Fermi surface (SFS) in these systems - together with the main FS (MFS) forming
the two primary features of the intrinsic fermiology of Bi2212 - which has eluded a definitive interpretation up to the
present.

In this letter, we prove that the SFS in Bi2212 and Pb-doped Bi2212 are intrinsic parts of the occupied electronic
states of these systems. By combining new data on the polarization dependence of these states in photoemission with
electron diffraction data and one-step {\it ab initio} photoemission simulations, we can definitively identify
structural distortions from tetragonality that result in the electronic bands giving rise to the SFS. Consequently, the
shadow Fermi surface in Bi2212 has lost its enigmatic character and is now needs a new name.

The SFS resembles a copy of the main FS, but shifted by ($\pi,\pi$) \cite{k-space-notation} and equivalent vectors in
the 2D tetragonal Brillouin zone, and was first observed in 1994 \cite{aebi-prl}. This feature is an accepted part of
all high quality ARPES data from Bi2212 and the related Bi2201 (Bi$_2$Sr$_2$CuO$_{6+\delta}$) high $T_c$
superconductors. Two groups of mechanisms have been proposed for the SFS. In the first, the shadows are magnetic: short
range antiferromagnetic (AFM) fluctuations being the culprit \cite{sfs-afm}. In the second, the shadows are of a
non-magnetic origin \cite{sfs-nonmag}, including the suggestion of back-folding of bands due to departure from the
ideal tetragonal structure of Bi2212 and related systems \cite{singh-prb}. Within the AFM scenario, significant
renormalization of the shadow band (SB) dispersion with respect to that of the main band (MB) should occur, and the SB
intensity should be strongly doping dependent. A recent ARPES study \cite{koitzsch-prb} argued convincingly against the
AFM scenario, due to the lack of additional renormalization, doping or temperature dependence of the SB signal in
photoemission. The question then arises: rather than simply stating that the AFM scenario is incorrect, can one  prove
that these primal Fermi surface features in the drisophila of high $T_c$ ARPES investigations are of structural origin?
Are they thus anchored in the initial states relevant not only for photoemission, but also for transport and other
physical properties?

In order to answer these questions, we carried out ARPES at the SIS beamline of the Swiss Light Source, with typical
energy and angular resolution of 34(40) meV and 0.3$^\circ$ for $h\nu$ = 40(100) eV. High quality pristine and Pb-doped
Bi2212 untwinned single crystals were grown using floating zone and flux methods, mounted on a 5-axis cryomanipulator
allowing polar and azimuthal rotation of the sample in ultra-high vacuum with a precision of 0.1$^\circ$. Cleavage and
measurement took place at T=15K.

In the geometry used, the synchrotron radiation, sample surface normal and the detected photoelectrons all lie in a
horizontal plane. The beamline provides linearly polarized radiation (polarized either horizontally [{\it
p}-polarization] or vertically [{\it s}-polarization]) or circularly polarized radiation (denoted $\sigma^+$ for left
circular polarization). As a result, for a high symmetry line such as (0,0)-($\pi$,$\pi$), which is a structural mirror
plane in the tetragonal cell \cite{k-space-notation}, we can classify the initial states as being either even or odd,
and thus expect to detect intensity from an even(odd) initial state only for {\it p}({\it s})-polarization. In this
context, $\sigma^+$ polarized light can be considered as a superposition of {\it p}- and {\it s}-polarization, and
consequently we use it to get a view of all features concerned.

\begin{figure}
\begin{center}
\includegraphics[width=3in]{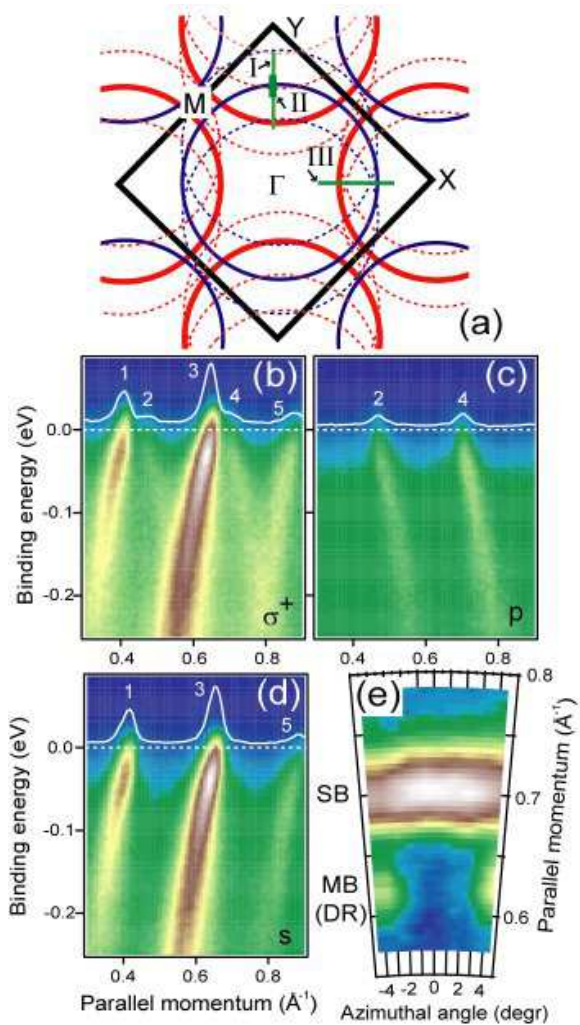}
\caption{\label{fig:gy-pristine} (color online) (a) Sketch of the Bi2212 Fermi surface. EDMs ($h\nu$ = 100 eV)
measured at location I in (a) using the polarizations indicated. The white lines show MDCs at $E_F$. (e) map of
intensity at $E_F\pm$15 meV from region II, measured with {\it s}-polarization and $h\nu=$40 eV.}
\end{center}
\end{figure}

Fig.~\ref{fig:gy-pristine}(a) shows a sketch of the Fermi surface of pristine Bi2212, including the MFS (red
barrels centered at X and Y \cite{k-space-notation}), the SFS (blue barrel centered at $\Gamma$) and their
diffraction replicas (DRs, dashed), the latter being caused by the incommensurate modulation of the BiO planes
running along the crystallographic $\mathbf{b}$-axis \cite{zandbergen-phc}. Panel (c) shows an energy distribution
map (EDM) for $\mathbf{k}$ along cut I ($\Gamma$Y), measured with $\sigma^+$ polarized light. Guided by
Fig.~\ref{fig:gy-pristine}(a), the features can be identified as: (1) the MB, (2) 1st order SB DR, (3) 1st order
MB DR, (4) SB and (5) 2nd order MB DR. Panel (d) shows the same cut, now measured with {\it p}-polarization. As
expected from the odd character of the Zhang-Rice singlet (ZRS) states \cite{zhang-prb} with respect to the
(0,0)-($\pi$,$\pi$) mirror plane, the MB features are completely suppressed for {\it p}-polarization.
Surprisingly, the shadow features are essentially unchanged with respect to the $\sigma^+$ case. Switching to {\it
s}-polarization (Fig.~\ref{fig:gy-pristine}[e]), we now see that the situation has reversed: only the MB related
features are visible, whereas the SB emission is suppressed. Fig.~\ref{fig:gy-pristine}(b) illustrates that the
'switching off' of the main band (in fact a DR of the main band, labelled 3) occurs only within $\pm2^\circ$ of
$\Gamma$Y, and also that the SB intensity is maximal exactly along the nodal line. The fact that the MB initial
states are odd with respect to the $\Gamma$Y line, and - unexpectedly - the SB initial states are {\it even} will
prove very important for the elucidation of the microscopic origin of the SFS.

\begin{figure}
\begin{center}
\includegraphics[width=2.5in]{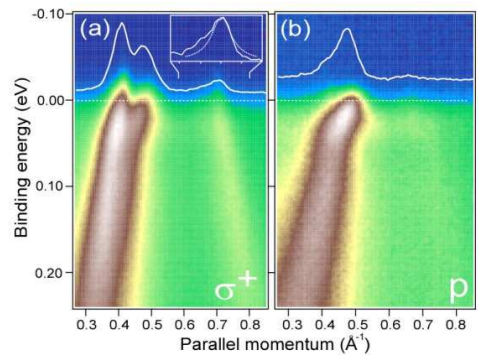}
\caption{\label{fig:gx-pristine} (color online) EDMs ($h\nu$ = 40 eV) measured at location III in
Fig.~\ref{fig:gy-pristine}(a), using the polarisations shown. The $E_F$ MDCs are shown as white lines. The inset
to panel (a) shows a blow-up of the $E_F$ MDC, including a Lorentzian as a guide to the eye.}
\end{center}
\end{figure}

We now turn to the $\Gamma$X direction, in which (for pristine Bi2212) the contributions from the main, shadow and
replica features are less widely spaced. In fact, the MB and its associated DRs are only separated by 0.07 \AA$^{-1}$,
which is resolvable in Fig.~\ref{fig:gx-pristine} by using the lower photon energy of 40 eV.
Fig.~\ref{fig:gx-pristine}(a) is recorded with $\sigma^+$ radiation. On the left is an intense feature related to the
MB, and on the right, a weak feature originating from the SB. At $E_F$, the momentum distribution curve (MDC) shows
that the left-hand feature can be resolved into two distinct branches, with that closest to $\Gamma$ being the main
band and the other being due to main band DRs (see also Fig.~\ref{fig:gy-pristine}[a]). After zooming the scale of the
MDC (see inset), a similar analysis is possible for the shadow states. Here the component furthest from $\Gamma$ is the
SB. The intensity observed in the inset at the left hand side of the Lorentzian is attributed to the SB DRs.

Fig.~\ref{fig:gx-pristine}(b) now shows the same $\Gamma$X cut, but measured with {\it p}-polarization. Now the
intensity from both the MB and the SB is highly suppressed, whereas the DRs of both bands are still observed. This
means that along $\Gamma$X, the MB and SB show {\it identical} polarization behavior, and originate from initial states
which are odd with respect to reflection in $\Gamma$X.

\begin{figure}
\begin{center}
\includegraphics[width=3in]{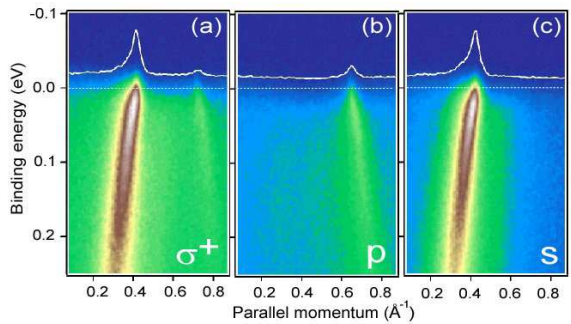}
\caption{\label{fig:gy-pbdoped} (color online) EDMs ($h\nu$ = 100 eV) from modulation-free (Bi,Pb)-2212 measured
at location I in Fig.~\ref{fig:gy-pristine}(a) using the three polarizations shown. The white lines show $E_F$
MDCs.}
\end{center}
\end{figure}

As ever in the Bi-based high $T_c$ superconductors, it is wise to check the validity of an important result in
modulation-free, Pb-doped crystals. Fig.~\ref{fig:gy-pbdoped}(a-c) shows the results for $\Gamma$Y. Using $\sigma^+$
polarization, we see both the MB and SB. For {\it p}[{\it s}] polarization (panels b[c]) the same behavior as in the
pristine case is evident with {\it opposite} polarization behavior of the MB (odd initial state) and SB (even initial
state).

By now the alert reader will pose the question: why is the symmetry character of the MB and SB alike in one nodal
direction, whereas in the orthogonal nodal direction they behave as opposites? The first point here is that the
initial state symmetry has to be involved, as for these data recorded along high symmetry lines we are able to
rule out an origin arising purely from multiple scattering in the photoemission final states\cite{strocov-prb}. As
a second point we recall that the AFM scenario for SFS formation has essentially been disqualified both from
earlier data\cite{koitzsch-prb} as well as from a similar analysis of the data presented here. Thus, our next step
is the suggestion that an orthorhombic structural distortion could lie at the root of things. Not only would this
lead to a back-folding of the MB into a smaller, orthorhombic Brillouin zone\cite{singh-prb}, there is also
experimental evidence for such a structural modification from low energy electron diffraction (LEED)
\cite{strocov-prb}, x-ray and neutron diffraction of both Bi2212 and (Pb,Bi)-2212 \cite{miles-phc, torardi-prb,
leligny-prb, flukiger-prb}. The trick in the tail is the explanation as to how this would lead to a change in
parity of the back-folded (SB) states, but only along {\it one} of the two nodal directions present in the basic
tetragonal structure.

To facilitate the discussion, we introduce a toy model in order to illustrate the qualitative behavior. We start from a
c(2$\times$2) unit cell in the undistorted tetragonal system, which would be, in-fact, non-primitive and which contains
two identical Cu atoms per plane, delivering two identical ZRSs. These states all have odd parity with respect to both
$\Gamma$Y and $\Gamma$X and give the MB: this tetragonal structure has no SB. If we now introduce a real orthorhombic
distortion \cite{ortho-distortion}, a SB appears. In this sense, the wave function responsible for the SB is the
\textit{difference} between the distorted (orthorhombic) and undistorted (tetragonal) wave functions. We can write the
overall wave function of the distorted system as $\phi_D(x,y)$ as: $\phi_D(x,y)= D(x)\phi(x,y)$, where $D(x)$ is the
distortion (along one direction in space)\cite{distortion-footnote} and $\phi(x,y)$ the undistorted wave function. If
the distortion is small, we can write: $D(x) = 1 + xD'(0) + \ldots$~~. The total wave function thus becomes
$\phi_D(x,y) = \phi(x,y) + xD'(0)\phi(x,y) + \ldots$~~, now reformulated in terms of the undistorted wave function plus
a distortion of strength $D'(0)$. The distortion - i.e. the feature responsible for the SB wave function - is
proportional to $x\phi(x,y)$. Crucially, since both the functions $x$ and $\phi$ are odd in $x$, their product will be
even. We note that a second order term (or other even numbered terms) in the perturbation would not lead to a parity
swap: only the odd terms can do this. This situation is reminiscent of the k-dependent parity swaps observed for states
in other systems possessing glide-mirror planes, such as pyrolytic graphite \cite{pescia-ssc} and C on Ni(100)
\cite{prince-ssc}.

The toy model, in fact, describes the essence of the surprising and new experimental results presented here. As evident
from diffraction results \cite{miles-phc, torardi-prb, leligny-prb, flukiger-prb}, the distorted structure only has a
single mirror plane, parallel to the xz-plane: the yz-plane being rather a glide-mirror plane (however, both the xz-
and yz-planes are mirror planes of the 'distortion wave function'). The XRD results indicate a significant shift of Bi
atoms in the BiO layers (by as much as 0.1 \AA) in the $x$ direction away from their ideal tetragonal sites, resulting
in two Bi sub-lattices. This shift is sketched in Fig. 4(a). Such displacements should and do show up as c(2$\times$2)
superstructure spots in LEED data \cite{strocov-prb}. We now take the LEED analysis an important step further with the
data shown in Fig. 4(b). The spots are indexed according to the {\it orthorhombic} $\mathbf{a}$ and $\mathbf{b}$ axes.
Clear superstructure spots (with respect to the tetragonal cell) are visible, as expected. A closer inspection reveals
that the ($h$=odd,$k$=0) spots are missing (independent of the electron beam energy). These systematic extinctions
point unambiguously to the presence of an orthorhombic distortion giving only a single nodal mirror plane running
parallel to xz, wholly in keeping with our simple model and Fig. 4(a).

\begin{figure}
\begin{center}
\includegraphics[width=3in]{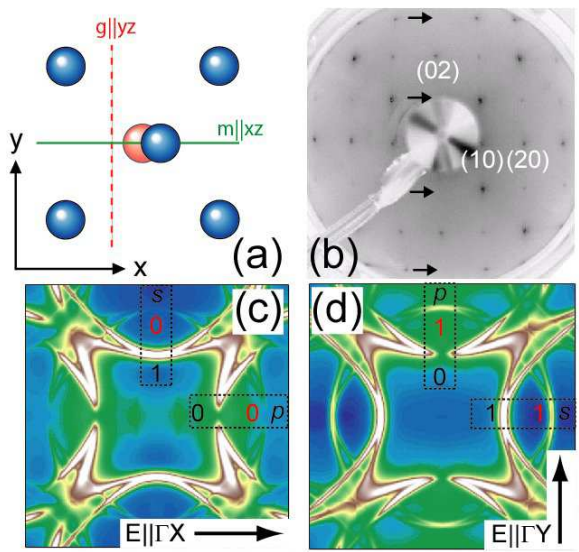}
\caption{\label{fig:leed-dft} (color online) (a) Shift of one Bi sub-lattice in the BiO planes along the x-axis,
whereby the red atom marks the undistorted position. m$\parallel$xz and g$\parallel$yz mark the mirror and
glide-mirror planes, respectively. (b) LEED pattern of (Bi,Pb)2212, systematic extinctions are indicated by
arrows. (c-d) Fermi surfaces calculated for 33 eV photons with the polarization vector aligned as indicated. The
dotted boxes indicate the relevant polarization and the black[red] '1's and '0's mark the presence and suppression
of the main[shadow] features, respectively.}
\end{center}
\end{figure}

In the final leg of our journey out of the shadows, we test the illustrative picture developed thus far by means
of DFT-based one-step photoemission calculations (for the methodology see Ref.~[\onlinecite{lindroos-prb}]). Here
we are able to deal explicitly with the \textbf{k}- and $\omega$-dependence of the matrix elements governing the
photoemission intensity. The input of the calculation is modulation-free Bi2212 in the average structure from
Ref.~\cite{miles-phc}. Figs. 4(c) and (d) show portions of calculated Fermi surface maps for polarization vectors
along x and y, respectively. From the calculated intensity distributions, we can see that the MFS displays the
expected behavior. It is suppressed along the zone diagonals for {\it p}-polarization and is strong for {\it
s}-polarization. Crucially, the SFS behaves differently along $\Gamma$Y, showing a parity swap along the vertical
nodal line. For the $\Gamma$X azimuth, the SFS follows the intensity distribution of the MFS. Thus, these {\it ab
initio} results for the realistic, distorted structure clearly show that the observed parity alteration for the
SFS is also an intrinsic part of the full photoionization matrix element. A more detailed group theoretical
analysis of these matrix element effects is currently underway, but goes beyond the scope of this letter.

Thus, in this last section of the paper, we have closed the
self-consistency loop between the anomalous polarization
dependence seen in ARPES, the extinctions seen in LEED and the
one-step photoemission calculations. All in all, this provides
watertight proof that the SFS in the Bi2212 and by (structural)
analogy also Bi2201-based high $T_c$ superconductors is an
intrinsic feature in the initial state electronic structure of
these much studied systems, arising from bulk, orthorhombic
distortions located primarily in the BiO planes, but most
definitely felt throughout the three-dimensional crystal. These
distortions appear neither to favour nor disfavour the
superconductivity itself. The fact that the most commonly adopted
experimental geometry for nodal ARPES experiments for these
systems is $\Gamma$Y with {\it s}-polarized radiation (i.e. a
geometry in which the SFS is invisible) helps explain the
imbalance between the number of experimental and theoretical
papers dealing with the SFS. Furthermore, the proof provided here
that orthorhombic distortions not only exist in the BiO layers of
pristine {\it and} Pb-doped Bi2212, but also that they have an
immediate impact on the CuO$_2$ plane states, is also highly
relevant to the controversy surrounding possible spontaneous
breaking of time reversal symmetry (TRS) in underdoped Bi-based
cuprate superconductors\cite{varma-prb-prl}. From the present data
it is now clear that the $\Gamma$M mirror planes relevant for
these TRS studies in Bi2212 \cite{TRS-exps} are not felt by the
CuO$_2$-plane states to be mirror planes at all - even in
modulation-free samples - thus disqualifying their use in attempts
to prove broken TRS. A further interesting consequence of the fact
that the SFS is rooted firmly in the initial states is that the
Bi-based HTSC therefore possess a fully nested FS, with every
point on the MFS being linked to an analogous point on the SFS by
the nesting vector ($\pi$,$\pi$).

In summary, by proving the microscopic structural origins of the shadow bands in the Bi2212 and Bi2201 families of
cuprate superconductors, we have finally been able to close this chapter in the rich and complex tale of the high $T_c$
superconductors.

We are grateful to W. Koops and T.J. Gortenmulder for technical support, to FOM (ILP and SICM) and the EU (I3) for
financial support and to A.A. Kordyuk and D.J. Singh for stimulating discussions. H.B. is grateful to the Fonds
National Suisse de la Recherche Scientifique and V.A. to the Finnish Academy and the Institute of Advanced Computing,
Tampere.


\begin{thebibliography}{}
\bibitem{k-space-notation} For simplicity, and to connect to existing literature on Bi2212, we have adopted the labels of the
tetragonal 2D Brillouin zone in which, for example, ($\pi$,$\pi$) is shorthand for ($\frac{\pi}{a}$,$\frac{\pi}{a}$)
with $a = 3.8$ \AA.
\bibitem{ortho-distortion} Here we mean more than merely setting the crystallographic a and b axes to be non-equal, but rather the adoption of a basic structure containing two inequivalent units per cell.
\bibitem{distortion-footnote} For simplicity, the y-dependence of D is omitted. As the xz plane is a mirror plane of both the overall distorted structure and the 'distortion wave function', the y-dependence of D cannot be of relevance for explaining the observed partity swap along $\Gamma$X.
\end{thebibliography}
\end{document}